\documentclass[pra,preprint,amsfonts,amssymb,amsmath,titlepage,floatfix,showpacs,superscriptaddress]{revtex4} 

\usepackage{psfrag}
\usepackage{bm}
\usepackage{verbatim}
\usepackage{bm}
\usepackage{graphicx}  
\usepackage{latexsym}                
\usepackage{amsmath}     
\usepackage{amsfonts}  
\usepackage{amssymb}
\usepackage{amsthm}
\usepackage{dcolumn}
\usepackage{color}

\newcommand{\be}{\begin{equation}}
\newcommand{\ee}{\end{equation}}
\newcommand{\bea}{\begin{eqnarray}}
\newcommand{\eea}{\end{eqnarray}}

\def\pder#1#2{\frac{\partial #1}{\partial #2}}
\def\softt{{\leavevmode\setbox1=\hbox{t}%
\hbox to \wd1{t\kern-0.6ex{\char039}\hss}}}

\begin{document}
\title{Early Stage of Superradiance from Bose-Einstein Condensates}

\date{\today}
\author{L.~F.~Buchmann}
\affiliation{Department of Physics, University of Crete, P.O. Box 2208, Herakleion 710 03, Crete, Greece}
\affiliation{Institute of Electronic Structure and Laser, Foundation of Research and Technology Hellas, P.O. Box 1527, Herakleion 711 10, Crete, Greece}
\author{G.~M.~Nikolopoulos}
\affiliation{Institute of Electronic Structure and Laser, Foundation of Research and Technology Hellas, P.O. Box 1527, Herakleion 711 10, Crete, Greece}
\author{O.~Zobay}
\affiliation{School of Mathematics, University of Bristol, University Walk, Bristol BS8 1TW, UK}
\author{P.~Lambropoulos}
\affiliation{Institute of Electronic Structure and Laser, Foundation of Research and Technology Hellas, P.O. Box 1527, Herakleion 711 10, Crete, Greece}
\affiliation{Department of Physics, University of Crete, P.O. Box 2208, Herakleion 710 03, Crete, Greece}

\begin{abstract}

We investigate the dynamics of matter and optical waves at the early stage of superradiant Rayleigh scattering 
from Bose-Einstein Condensates. Our analysis is within a spatially dependent quantum model which is capable of 
providing analytic solutions for the operators of interest. The predictions of the present model are compared to 
the predictions of a closely related mean field model, and we provide 
a procedure that allows one to calculate quantum expectation values by averaging over semiclassical solutions. 
The coherence properties of the outgoing scattered light are also analyzed, and it is shown that the 
corresponding correlation functions may provide detailed information about the internal dynamics of the system.   
\end{abstract}

\maketitle
\section{Introduction}

Superradiance in general terms refers to enhanced emission from an ensemble of radiators. 
It was first predicted by Dicke in 1954 \cite{Dicke54} and since then experimentally confirmed to occur in many systems, 
such as gases of excited atoms, molecules or quantum dots \cite{SR,GroHar82}. 
Recently, superradiant scattering off an elongated atomic Bose-Einstein condensate (BEC) has received much 
theoretical \cite{SRTheorSemiclass,SRTheorQuant,ZobNikPRA,MooMeyPRL} and experimental \cite{SRExp,InoChiSta99,VogXuKett} attention. 
There are many similarities, but also important differences between the ``conventional'' 
superradiance, for example off excited gases, and the superradiance off atomic condensates. 
Atoms in a BEC have a narrow momentum distribution, and thus the recoil they experience during the 
absorption and emission of photons has a profound impact on their momentum distribution, leading 
to distinct atomic scattering patterns.

In the case of superradiant Rayleigh scattering off BECs, different regimes of parameters have been identified, 
which are characterized by distinct atomic patterns. 
Mean-field models were found to successfully predict and explain such patterns, as well as the transition between 
different regimes, provided the models include spatial effects \cite{SRTheorSemiclass,ZobNikPRA}. The main drawback of such models, however, 
is that one has to ``seed'' the equations of motion, in order to start their evolution in time.  
The seeding introduces some ambiguity in the solutions, which is expected to become less important 
for large times due to the fast growth of the population in the various optical and mater-wave modes.  
In contrast to the mean-field models, the quantum models that have been used in this context, are capable of 
describing accurately the startup of the process, but do not take into account 
spatial propagation effects \cite{SRTheorQuant,MooMeyPRL}. 

In a recent work \cite{BuNikZobLamRC}, we investigated the coherence properties of matter waves produced in 
superradiant scattering off BECs, and analyzed the type of spatial correlations involved. 
This has been possible in the framework of a spatially-dependent quantum model, which can describe 
quantum fluctuations while capturing spatial effects, essential for a full understanding of the process.  
The purpose of the present paper is to provide a full derivation of the model used in \cite{BuNikZobLamRC}, 
and to obtain further detailed insights into the dynamics of the system. We explicitly compare our results to 
those obtained within a related mean-field model, and show that for a large collection of condensed atoms, 
the effect of quantum fluctuations on various observable quantities can be obtained by averaging over 
ensembles of semiclassical solutions (``trajectories''). Finally, we discuss the temporal coherence of the 
scattered light, which is shown to carry information about the internal dynamics of the system.

\section{The Model}
 
\label{secII}
The system under consideration pertains to a BEC elongated along the $z$-axis, consisting of $N$ atoms. 
A linearly polarized laser with frequency $\omega_l=k_l c$, far detuned from the closest atomic transition by a value of $\delta$, is illuminating the cloud along the $x$-axis. 
Assuming two-level atoms and adiabatically eliminating the excited atomic state, the Maxwell-Schr\"odinger equations for the system read \cite{GroHar82,ZhangWalls}
\begin{subequations}\label{maxschrod}
\begin{align}
{\mathrm i}\hbar\pder{}t\hat\psi({\bf x},t) &= -\frac{\hbar^2}{2M}\Delta\hat\psi({\bf x},t)\nonumber\\
&\quad +\frac{[{\bf d}\cdot\hat{\bf E}^{(-)}({\bf x},t)][{\bf d}\cdot\hat{\bf E}^{(+)}({\bf x},t)]}{\hbar\delta} \hat\psi({\bf x},t),\label{maxschrodpsi}\\
\pder{^2 \hat{\bf E}^{(\pm)}({\bf x},t)}{t^2} &= c^2 \Delta \hat{\bf E}^{(\pm)}({\bf x},t)-\frac{1}{\varepsilon_0} \pder{^2 \hat{\bf P}^{(\mp)}({\bf x},t)}{t^2},\label{maxschrodE}
\end{align}
with the atomic polarization
\be
\label{pol_ad}
\hat{\bf P}^{(-)}({\bf x},t) = -{\bf d} \hat\psi^\dagger({\bf x},t) \frac{{\bf d}\cdot\hat{\bf E}^{(+)}({\bf x},t)}{\hbar\delta} \hat\psi({\bf x},t),
\ee
\end{subequations}
satisfying ${\bf P}^{(+)}={{\bf P}^{(-)}}^\dag$.
The operators $\hat{{\bf E}}^{(\pm)}({\bf x},t)$ are the positive and negative frequency parts of the electromagnetic field, while $\hat{\psi}({\bf x},t)$ is the operator describing the ground state of the atoms, which have mass $M$ and dipole moment ${\bf d}$. In writing Eqs. (\ref{maxschrod}) we have neglected the external trapping potential as well as atomic interactions, which both do not play a significant role for the timescales of interest. 
Due to the coherent nature of the condensate, successive Rayleigh scattering events are strongly correlated and lead to collective superradiant behavior. 
As a result of the cigar-shape of the condensate, the gain is largest when the scattered photons leave the condensate along its long axis, traveling in the so called {\it endfire modes} with wave vectors ${\bf k}\approx \pm k_l{\bf e}_z$.
A condensate atom can scatter a laser photon into the endfire modes, experiencing a recoil $\hbar{\bf q}\approx\hbar(k_l{\bf e}_x -{\bf k})$. On the other hand, it can also scatter a photon from the endfire modes into the laser mode, in which case its momentum changes by $\hbar{\bf q}\approx \hbar (-k_l{\bf e}_x + {\bf k})$. These processes lead to the formation of two pairs of atomic {\it side-modes}, consisting of counterpropagating atoms with a narrow momentum spread (compared to $k_l$). 
Of course, atoms within these side-modes can also scatter photons, thereby acquiring higher momenta, but since we are interested in the early stage of the process, we consider only first order sidemodes to be populated.

Neglecting any coupling {\it between} counterpropagating photonic endfire modes, 
the system becomes symmetric with respect to the $x$-axis. We can thus focus on the endfire modes 
with ${\bf k}\approx +k{\bf e}_z$ and on the two atomic side-modes 
(with central momenta $\hbar{\bf q}=\pm\hbar(k_l{\bf e}_x - k {\bf e}_z)$) 
that are coupled to them. Due to the strong confinement along the $x$- and $y$-axis, 
we can assume the transverse profiles of  the matter field $\psi_\perp(x,y)$  to be well 
described by a classical function, independent of the $z$-coordinate \cite{ZobNikPRA}. 
Assuming a Fresnel number close to unity for the electromagnetic fields, we can apply the 
same approximation for the transverse part of the radiation field $u_\perp(x,y)$, effectively 
reducing the problem to one dimension.

We expand the field operators as
\begin{subequations}\label{expansions}
\begin{align}
\hat{\psi}({\bf x},t)&=\psi_\perp(x,y)\sum_{j=-1}^1\hat{\psi}_{j}(z,t)e^{{\mathrm i}j(k_lx-kz)-{\textrm i}\omega_jt}\label{expandpsi}\\
\label{expandE}
\hat{\bf E}^{(+)}({\bf x},t) &=\frac{{\mathcal E}_0}{2}{\bf e}_y e^{{\mathrm i}(k_lx-\omega_l t)} + u_\perp(x,y)\hat{E}_+^{(+)}(z,t){\bf e}_y,
\end{align}
\end{subequations}
where $\omega_{\pm 1}=\hbar(k_l^2+k^2)/2M$ and $\omega_0=0$. The matter-wave operator is split up in three parts, describing the two side-modes ($j=\pm 1$) and the BEC at rest ($j=0$), 
\begin{subequations}\label{modeexp}
\be
\hat{\psi}_{j}(z,t)=e^{{\mathrm i}\omega_{j}t}\sum_{p\in\Delta_{0}}\frac{e^{\mathrm i p z}}{\sqrt{L}} {\hat c}_{-j k+p}(t),\label{psipm}
\ee
where $\Delta_0$ is the interval $(-k/2,k/2)$ in $k$-space, $L$ is the length of the BEC and $\hat{c}_p$ annihilates an atom with momentum $\hbar p$. Since the BEC at rest remains practically undepleted it can be treated as a time independent classical function and hence we can set $\hat{\psi}_{0}(z,t)\equiv\psi_0(z)$. Similarly, we expand the endfire mode operator as
\bea
\hat{E}_+^{(+)}({\bf z},t) &=& \mathrm i e^{\mathrm i(kz-\omega t)} \sum_{p\in\Delta_{0}} \sqrt{\frac{\hbar \omega_{k+p}}{2\varepsilon_0 }}  \frac 1{\sqrt L} e^{\mathrm ipz}e^{\mathrm i\omega t} \hat a_{k+p}(t)\nonumber\\
&\approx& \sqrt{\frac{\hbar \omega}{2\varepsilon_0}}e^{\mathrm i(kz-\omega t)}\hat e_+(z,t)\label{exp_Ep},
\eea
\end{subequations}
where $\hat{a}_p$ is the photon annihilation operator. The frequencies $\omega_{k+p}$ are approximated by $\omega=k/c$, the frequency of the scattered photons, and can therefore be taken out of the sum. This approximation is justified by the fact that dominant contributions to the sum come from momenta of order $1/L$, which is several orders of magnitude smaller than $k$.

The one-dimensional field operators satisfy the commutation relations
\begin{subequations}\label{commrel}
\bea
\left[\hat{\psi}_i(z_1,t),\hat{\psi}_j^\dag(z_2,t)\right]&=&\delta_{ij}\delta_\Delta(z_1-z_2),\label{commrelpsi}\\
\left[\hat{e}_{+}(z_1,t),\hat{e}_{+}^\dag(z_2,t)\right]&=&\delta_\Delta(z_1-z_2),\label{commrelExi}\\
\left[\hat{e}_{+}(z,t_1),\hat{e}_{+}^\dag(z,t_2)\right]&=&\frac{1}{c}\delta_\Delta(t_1-t_2),\label{commrelEtau}
\eea
\end{subequations}
where $\delta_{\Delta}(z)$ is a distribution with width of order $1/k$ and $\delta_{ij}$ denotes the Kronecker delta \footnote{Since the fields of interest vary slowly on length scales $1/k$, we can approximate the $\delta_{\Delta}(z)$ distributions by Dirac delta functions}.
Inserting expansions (\ref{expansions}) in the equations of motion (\ref{maxschrod}), we can make some further simplifications. Considering that $k_l$ is about a factor of 10 smaller than the extent of the BEC cloud along the strongly confined axes, the two transverse functions $\psi_\perp(x,y)e^{\pm \mathrm i k_l x}$ are mutually orthogonal to a very good degree of approximation. Hence we can project on either of the two side-mode operators by multiplying with the complex conjugate of the corresponding transverse function and integrate over the variables $x$ and $y$.

Since we included the phase factors arising from the free time evolution in the definition of the operators, we can apply the slowly-varying-envelope approximation (SVEA), which yields the equations
\begin{subequations}\label{eom}
\begin{align}
\frac{\partial \hat{\psi}_{+1}^\dag(\xi,\tau)}{\partial\tau}&=
{\rm i}\kappa \hat{e}_+(\xi,\tau) \psi_{0}^*(\xi)
,\label{eom1-1}\\
\frac{\partial \hat{\psi}_{-1}^\prime(\xi,\tau)}{\partial\tau}&=
-{\rm i}\kappa \hat{e}_+(\xi,\tau)\psi_{0}(\xi) 
-2{\rm i} \hat{\psi}_{-1}^\prime(\xi,\tau)
,\label{eom-11}\\
\frac{\partial \hat{e}_+(\xi,\tau)}{\partial\tau}+
\chi \frac{\partial \hat{e}_+(\xi,\tau)}{\partial\xi}
&=-{\rm i}\left [\kappa\psi_{0}(\xi)
\hat{\psi}_{+1}^\dag(\xi,\tau)\right.\nonumber\\
&\left.\quad+
\kappa\hat{\psi}_{-1}^\prime(\xi,\tau)
\psi_{0}^*(\xi)\right ].
\label{eome+}
\end{align}
\end{subequations}
Here we have defined $\hat{\psi}_{-1}^\prime(\xi,\tau)=e^{-2{\mathrm i}\tau}\hat{\psi}_{-1}(\xi,\tau)$ and rescaled length  and time to dimensionless units 
\bea
\xi=k_lz, \qquad \tau=2\omega_rt,
\eea
where $\omega_r=\hbar k_l^2/2M$. Accordingly, the fields are rescaled as
\be
\hat{e}_+(\xi,\tau)\equiv\frac{1}{\sqrt{k_l}}\hat{e}_+(z,t)\qquad\hat{\psi}_{j}(\xi,\tau)\equiv\frac{1}{\sqrt{k_l}}\hat{\psi}_{j}(z,t)\nonumber,
\ee
and the speed of light becomes $\chi\equiv\frac{ck}{2\omega_r}$.
The effective one-dimensional coupling is given by $\kappa=g\sqrt{k_lL}/(2\omega_r)$ with
\be
g=\frac{|{\bf d}\cdot{\bf e}_y|^2{\mathcal E}_0}{\hbar^2\delta}\sqrt{\frac{\hbar\omega}{2\epsilon_0L}}\int{\mathrm d}x{\mathrm d}yu_\perp(x,y)\psi_\perp^2(x,y).\nonumber
\ee
The SVEA pertained to neglecting derivatives of slowly varying functions in order to arrive at Eqs. (\ref{eom1-1}), (\ref{eom-11}). For Eq. (\ref{eome+}), we have kept first order derivatives, but neglected terms proportional to $\hat{e}_+(z,t)$ in comparison to the laser field. The frequencies in our system satisfy  $\omega_{\pm 1}\ll \omega_l,\omega$ and thus we only kept time derivatives involving $e^{- \mathrm i\omega_l t}$ and approximated $\omega\approx\omega_l$.

Backwards recoiling atoms are a particular feature of superradiant Rayleigh scattering off condensates. 
The physical process underlying the backwards modes violates energy conservation by an amount 
$\Delta E\simeq 4 \hbar\omega_{r}$. Thus, according to Heisenberg uncertainty principle, it can take place 
only for times shorter than a critical time $t_{c}=\hbar/\Delta E$ \cite{InoChiSta99},
 which, in our units, is given by 
\be
\tau_c=0.5.
\ee 
For such short pulses, and for sufficiently high power one typically 
observes an X-shaped pattern for the distribution of the atomic side modes with the initial BEC in the center and the 
recoiling atoms moving both in and against the direction of the applied laser pulse 
({\it Kapitza-Dirac} or {\it strong-pulse regime}). On the other hand, 
for weaker pulses with duration longer than $\tau_{c}$, 
the distribution of the side modes exhibits a fan pattern, involving mainly forward recoiling atoms 
({\it Bragg} or {\it weak pulse regime}). 
If we neglect the atomic backwards sidemode altogether, Eqs. (\ref{eom}) become formally equivalent to descriptions of ``conventional'' superradiance from excited atomic gases \cite{Dicke54}. 
 
Finally, the equations of motion (\ref{eom}) can be derived from the effective, self-adjoint Hamiltonian 
\be
\hat H=\int\mathrm{d}\xi\left(-\frac{\mathrm{i}}{2}\hat e_+^\dag\frac{\partial\hat{e}_+}{\partial\xi}+\hat{\psi}_{-1}^{\prime\dag}\hat{\psi}_{-1}^{\prime}+\kappa\psi_0\hat{e}^\dag_+\hat{\psi}_{+1}^\dag+\kappa\psi_0^*\hat{e}_+^\dag\hat\psi_{-1}^\prime+\mathrm{h.c.}\right),
\ee
where ``h.c.'' stands for the Hermitian conjugate. The system being effectively hamiltonian 
guarantees conservation of the commutation relations (\ref{commrel}) for all times. 
Differentiating the atomic densities 
\begin{subequations}
\be
n_j(\xi,\tau)=\langle\hat\psi^\dag_{j}(\xi,\tau)\hat\psi_j(\xi,\tau)\rangle 
\label{ad}
\ee
and the photon density 
\be
\mathfrak{I}(\xi,\tau)=\langle\hat{e}_+^\dag(\xi,\tau)\hat{e}_+(\xi,\tau)\rangle 
\label{pd}
\ee
\end{subequations}
with respect to time, using Eqs. (\ref{eom}) and adding up the resulting three equations, 
we find the continuity equation
\be
\frac{\partial}{\partial\tau}\left[n_{+1}(\xi,\tau)-n_{-1}(\xi,\tau)-\mathfrak{I}(\xi,\tau)\right]=\chi\frac{\partial}{\partial\xi}\mathfrak{I}(\xi,\tau).
\ee
 Let us now integrate this equation over time from $0$ to $\tau$ and over space from one end of the condensate at $\xi=0$ to the other end at $\xi=\Lambda\equiv k_lL$. Assuming that side-mode and photon populations vanish at $\tau=0$, we find
\be
\mathcal N_{+1}(\tau)-\mathcal N_{-1}(\tau)-\mathcal I_{\mathrm{in}}(\tau)=\mathcal I_{\mathrm{out}}(\tau),
\label{cont_eq}
\ee 
where we have defined the total populations for the atoms
\begin{subequations}
\be
{\mathcal N}_j(\tau)=\int_0^\Lambda {\mathrm d}\xi n_j(\xi,\tau), \\ 
\label{totalpops}
\ee
and the photons
\bea
&&\mathcal I_{\mathrm{in}}(\tau)=\int_0^\Lambda \mathrm{d}\xi \mathfrak{I}(\xi,\tau),\\
\label{photonpop_in}
&& \mathcal I_{\mathrm{out}}(\tau)=\chi\int_0^\tau \mathrm{d}\tau' \mathfrak{I}(\Lambda,\tau'). 
\label{photonpop_out}
\eea
\end{subequations} 
In words, Eq. (\ref{cont_eq}) expresses that at any time $\tau$,  
the number of forward-recoiling atoms ${\mathcal N}_+(\tau)$, is equal to the sum 
of backwards recoiling atoms ${\mathcal N}_-(\tau)$, and endfire photons 
inside  and outside  the BEC volume, denoted by $\mathcal I_{\mathrm{in(out)}}(\tau)$. It is therefore consistent with the intuitive picture of the underlying process and it may serve as a convenient check for numerical simulations.

\section{Solutions of the Equations of Motion}
We can use the Laplace transform to find exact solutions to the system (\ref{eom}) in terms of the operators evaluated at the boundary of their domain -- i.e. at $\xi=0$ and $\tau>0$ or vice versa at $\xi>0$ and $\tau=0$. More details on this procedure are given in the appendix. The solutions read
\begin{widetext}
\begin{subequations}\label{solutions}
\begin{align}
\hat{e}_+(\xi,\tau)=&
\int_0^\tau d\tau^\prime \hat{e}_+(0,\tau^\prime ) F_{0,0}(\gamma_{\xi,0},\tau-\tau^\prime-\beta_{\xi,0})
-\frac{{\rm i}\kappa}{\chi}
\int_{0}^{\xi}d\xi^\prime
\left[
\psi_{0}(\xi^\prime)
\hat{\psi}_{+1}^\dag(\xi^\prime,0) F_{1,0}(\gamma_{\xi,\xi'},\tau-\beta_{\xi,\xi'})\right.\nonumber\\
&\left.+\psi_{0}^*(\xi^\prime)
\hat{\psi}_{-1}^\prime(\xi^\prime,0) F_{0,1}(\gamma_{\xi,\xi'},\tau-\beta_{\xi,\xi'})
+\frac{{\mathrm i}}{\kappa}\hat{e}_+(\xi^\prime,0)F_{0,0}(\gamma_{\xi,\xi'},\tau-\beta_{\xi,\xi'})\right ],
\label{e+1}
\\
\hat{\psi}_{+1}^\dag(\xi,\tau)=&
{\rm i} \kappa \psi_{0}^*(\xi)
\int_0^\tau d\tau^\prime \hat{e}_+(0,\tau^\prime ) F_{1,0}(\gamma_{\xi,0},\tau-\tau^\prime-\beta_{\xi,0})
+\hat{\psi}_{+1}^\dag(\xi,0)
\nonumber\\
&+\frac{\kappa^2}{\chi}\psi_{0}^*(\xi)
\int_{0}^{\xi}d\xi^\prime
\left [
\psi_{0}(\xi^\prime)
\hat{\psi}_{+1}^\dag(\xi^\prime,0) 
F_{2,0}(\gamma_{\xi,\xi'},\tau-\beta_{\xi,\xi'})
\right.\nonumber\\
&+\psi_{0}^*(\xi^\prime)
\hat{\psi}_{-1}^{\prime}(\xi^\prime,0)F_{1,1}(\gamma_{\xi,\xi'},\tau-\beta_{\xi,\xi'})
\left.+\frac{{\textrm i}}{\kappa}\hat{e}_+(\xi^\prime,0) F_{1,0}(\gamma_{\xi,\xi'},\tau-\beta_{\xi,\xi'})
\right ],
\label{psi_f1}
\\
\hat{\psi}_{-1}^\prime(\xi,\tau)
=&
-{\rm i}\kappa\psi_{0}(\xi)
\int_0^\tau d\tau^\prime \hat{e}_+(0,\tau^\prime ) F_{0,1}(\gamma_{\xi,0},\tau-\tau^\prime-\beta_{\xi,0})
+e^{-{\rm i}2\tau}\hat{\psi}_{-1}^\prime(\xi,0)
\nonumber\\
&-\frac{\kappa^2}{\chi}\psi_{0}(\xi)
\int_{0}^{\xi}d\xi^\prime
\left [
\psi_{0}(\xi^\prime)
\hat{\psi}_{+1}^\dag(\xi^\prime,0) F_{1,1}(\gamma_{\xi,\xi'},\tau-\beta_{\xi,\xi'})\right.\nonumber\\
&\left.+\psi_{0}^*(\xi^\prime)
\hat{\psi}_{-1}^\prime(\xi^\prime,0) F_{0,2}(\gamma_{\xi,\xi'},\tau-\beta_{\xi,\xi'})
+\frac{{\textrm i}}{\kappa}\hat{e}_+(\xi^\prime,0)F_{0,1}(\gamma_{\xi,\xi'},\tau-\beta_{\xi,\xi'})
\right ],
\label{psi_b1}
\end{align}
\end{subequations}
\end{widetext}
where we have introduced
\be
\beta_{\xi,\xi'}=\frac{\xi-\xi'}{\chi}, \quad \gamma_{\xi,\xi'}=\frac{\kappa^2}{\chi}[\rho(\xi)-\rho(\xi')]\nonumber,
\ee
with $\rho(\xi)=\int_0^\xi d\xi' |\psi_{0}(\xi')|^2$. The functions $F_{\mu,\nu}(u,v)$ are defined as
\be
F_{\mu,\nu}(u,v)={\mathcal L}_{p\to v}^{-1}\left\{\frac{e^{u/p}e^{-u/(p+2{\mathrm i})}}{p^\mu(p+2{\mathrm i})^\nu}\right\}\nonumber,
\ee
where ${\mathcal L}^{-1}_{p\to v}$ denotes the inverse Laplace transform. One can check easily that Eqs. (\ref{solutions}) indeed are solutions to the system (\ref{eom}) by using recursion relations for the functions $F_{\mu,\nu}$ which are given in the appendix, alongside the explicit expressions of the functions themselves.
Explicit expressions for the functions $F_{\mu,\nu}(u,v)$ appearing in (\ref{solutions}) are given in the appendix. They are combinations and integrals over combinations of Bessel Functions. It shall only be noted here, that all of the terms appearing in $F_{\mu,\nu}(u,v)$ contain Heaviside step functions $\Theta(v)$, except one term in $F_{0,0}(u,v)$, which is simply the Dirac delta function with argument $v$.

We note that in Eqs. (\ref{solutions}), all time arguments are shifted by the 
value $\beta_{\xi,\xi'}$, which is the time a photon needs to travel from $\xi'$ to $\xi$. 
Using the step 
functions in the solutions to change the range of the integrals and assuming free light 
propagation outside the condensate we can reformulate Eqs. (\ref{solutions}), such that 
they involve only spatial integrals ranging from $\xi-\tau\chi$ to $\xi$. 
This is a consequence of the finite speed of light, allowing atoms at $\xi$ only to be 
influenced by atoms within a range $\xi-\tau\chi$. Such retardation effects are very small 
in the system at hand and can be neglected for all practical purposes. We can do so 
formally by letting $\chi\to\infty$, which implies $\beta_{\xi,\xi'}\to 0$, and neglecting 
all the terms proportional to $\hat{e}_+(\xi',0)$ in the spatial integrals of Eqs. (\ref{solutions}).
The resulting solutions will still describe the system correctly, since the effects of this 
approximation are expected to be of the order of $\Lambda/\chi\approx10^{-7}$, and are thus too small 
to be noted in typical BEC experiments. Formally, this approximation will lead to a nonzero 
initial photon population within the BEC, which we can safely neglect due to its small value.  

Equations (\ref{solutions}) can be simplified considerably if we neglect backward recoiling atoms. 
Neglecting retardation effects, we obtain 
\begin{subequations}
\label{pfs}
\bea
\label{pfs-e}
\hat{e}_+(\xi,\tau)&=&
\int_0^\tau d\tau^\prime \hat{e}_+(0,\tau^\prime ) F_{0}(\gamma_{\xi,0},\tau-\tau^\prime)
\nonumber\\
&&-\frac{{\rm i}\kappa}{\chi}
\int_{0}^{\xi}d\xi^\prime
\psi_{0}(\xi^\prime)
\hat{\psi}_{+1}^\dag(\xi^\prime,0) F_{1}(\gamma_{\xi,\xi'},\tau)\\
\hat{\psi}_{+1}^\dag(\xi,\tau)&=&
{\rm i} \kappa \psi_{0}^*(\xi)
\int_0^\tau d\tau^\prime \hat{e}_+(0,\tau^\prime ) F_{1}(\gamma_{\xi,0},\tau-\tau^\prime)
+\hat{\psi}_{+1}^\dag(\xi,0)\nonumber\\
&&+\frac{\kappa^2}{\chi}\psi_{0}^*(\xi)
\int_{0}^{\xi}d\xi^\prime
\psi_{0}(\xi^\prime)
\hat{\psi}_{+1}^\dag(\xi^\prime,0) 
F_{2}(\gamma_{\xi,\xi'},\tau)
,
\label{pfs-psi}
\eea
\end{subequations}
where  $F_{\mu}(u,v)={\mathcal L}_{p\to v}^{-1}\left\{e^{u/p}p^{-\mu}\right\}$; explicit formulas for $F_\mu(u,v)$ are given in the appendix. It is worth emphasizing that these equations are consistent with the equations other authors derived to describe conventional superradiance \cite{Haake}. 

Assuming that the initial population of the atomic side modes is zero,  we can find the expectation value of any correlation function pertaining to electromagnetic-- or matter-wave fields using Eqs. (\ref{commrel}) and (\ref{solutions}) and calculating the occurring integrals  numerically.
For the numerical calculations, we assumed the BEC to consist of $N=10^6$ Thomas-Fermi distributed ${}^{87}\textrm{Rb}$ atoms, such that $\psi_0=\sqrt{\Theta(z)N6(Lz-z^2)/L^3}$ with $L=130\mu{\textrm m}$. We used a spatial grid of 400 points. For the incoming laser we chose a rectangular profile and a wavenumber $k_l=8.05\times10^6{\textrm m}^{-1}$, which results in a dimensionless length of the BEC of $\Lambda\sim 1000$. Coupling strengths are conveniently expressed in terms of the {\it superradiant gain} $\Gamma=\kappa^2N/\chi$, whose value separates the two regimes identified by experimental observations of superradiance from condensates 
\cite{VogXuKett}. Typically, the weak coupling regime is characterized by $g \sim 10^5{\textrm s}^{-1}$ and $\Gamma\sim1$, while for $g\sim10^6{\textrm s}^{-1}$ and $\Gamma\gg 1$ the system is in the strong coupling regime. In our calculations, we chose $\Gamma=1$ and $\Gamma=100$ for the two regimes.

\section{Quantum vs. Mean Field Description}
Various aspects of the strong and weak-coupling regimes have been described 
successfully within a mean-field (MF) model \cite{ZobNikPRA}, 
which is closely related to  the present quantum model given by Eqs. (\ref{eom}). 
In fact, one arrives at Eqs. (15) of \cite{ZobNikPRA} by adapting the 
approximations of the present quantum model and replacing the operators in Eqs. (\ref{eom}) 
with their expectation values, treating them as classical fields. 
Consequently, we will refer to the solutions of the mean field model as $\psi_j(\xi,\tau)\equiv\langle\hat\psi_j(\xi,\tau)\rangle$ and 
$e_+(\xi,\tau)\equiv\langle\hat e_+(\xi,\tau)\rangle$. Due to the generality of the Laplace transform, these solutions look exactly like Eqs. (\ref{solutions}), but with the operators replaced by classical fields.

Both models take into account spatial effects, which have been shown to play a major role in 
Rayleigh superradiance from condensates \cite{ZobNikPRA}. 
Given, however, that our system is initially prepared in the vacuum state, both 
 $\psi_j(\xi,\tau)$ and $e_+(\xi,\tau)$ will remain zero throughout the evolution of the system, 
because their equations of motion  [see Eqs. (\ref{solutions})], 
are not coupled to any operator with non-zero expectation value 
[i.e., $\psi_j(\xi,0) =0$ and  $e_+(\xi,0) =0$]. 
This is a major drawback of the MF model, which can be resolved by seeding either of the modes 
i.e., assigning a non-zero initial value to 
either $\psi_{j}(\xi,0)$ or  $e_+(\xi,0)$. The arbitrariness of such a seeding introduces 
some ambiguity regarding the dynamics of the system for short times, 
where all the modes are scarcely populated. The MF model is expected to be valid for 
longer times, where the fast growth of the population in the modes eliminates any ambiguity 
caused by the initial seeding. At such times, the MF model explains reasonably well various experimental observations \cite{ZobNikPRA}.
Our purpose in this section is to investigate how accurately one can describe the 
onset of superradiance from condensates, in the framework of this MF model. To this end, we will compare the predictions of the MF model for various 
observables, to the corresponding predictions of the quantum model, which 
is capable of describing accurately initial quantum mechanical fluctuations, and does not require 
any seeding.

\subsection{Atomic Densities and Populations}
A rather convenient observable in a superradiant scattering process from a BEC, is the atomic density of 
the sidemode $j$, denoted by $n_j(\xi,\tau)$.  
After turning off the atomic trap, atoms in the sidemodes separate from the BEC at rest due to their 
additional momentum and form observable scattering patterns \cite{InoChiSta99}. 

In the quantum model, we have $n_j(\xi,\tau)=\langle\hat\psi_j^\dag(\xi,\tau)\hat\psi_j(\xi,\tau)\rangle$, which in view of Eqs. (\ref{solutions}) yields  
\begin{subequations}\label{densities}
\begin{align}
n_{-1}(\xi,\tau)=&\Gamma^2|\varphi(\xi)|^2
\int_0^\xi{\mathrm d}\xi'|\varphi(\xi')|^2|F_{1,1}(\gamma_{\xi,\xi'},\tau)|^2,\label{backwg1}\\
n_{+1}(\xi,\tau)=&n_{-1}(\xi,\tau)+\Gamma|\varphi(\xi)|^2\int_0^\tau{\mathrm d}\tau'|F_{1,0}(\gamma_{\xi,0},\tau')|^2\label{forwg1},
\end{align}
\end{subequations}
where $\varphi(\xi)=\psi_0(\xi)/\sqrt{N}$. In the MF model, expectation values are defined as the squared modulus of the classical functions, i.e. $n_j(\xi,\tau) = |\psi_j(\xi,\tau)|^2$. By means of these quantities, we can directly compare the two models. 
\par
As a first step, let us neglect for the time being backward recoiling atoms. In this case 
one can obtain analytic expressions for the atomic densities, which acquire particularly 
simple forms for a flat BEC [i.e., for $\psi_{0}(\xi)=\sqrt{N/\Lambda}$].
Equation (\ref{forwg1}) reduces to 
\be
n_{+1}(\xi,\tau) = \bar{\Gamma}\tau\left [I_0^{2}(2\sqrt{\bar{\Gamma}\xi\tau})-I_1^{2}(2\sqrt{\bar{\Gamma}\xi\tau}) \right ],
\label{forwg1_pfs}
\ee
with $\bar{\Gamma}=\Gamma/\Lambda$, whereas within the MF model one obtains  
\be
n_{+1}(\xi,\tau) = \frac{|\psi_{+1}(\xi,0)|^2}{\Lambda}I_0^2(2\sqrt{\bar{\Gamma}\xi\tau}).
\ee
Note here that the prefactor of the Bessel functions in the case of the quantum model is a linear function of time, 
as opposed to the time-independent variable $|\psi_{+1}(\xi,0)|^2/\Lambda$ in the case of 
the MF model. This is a key difference, whose implications become clearer if we look at 
the asymptotic behavior of the total population of the forward atomic side mode. 
Using Eq. (\ref{forwg1_pfs}) in Eq. (\ref{totalpops}), we obtain for $\Gamma\tau\gg 1$ 
\be
{\cal N}_{+1}(\tau) \sim \frac{1}{16\pi\sqrt{\Gamma\tau}}e^{4\sqrt{\Gamma\tau}},
\ee
while for the MF model one finds
\be
{\cal N}_{+1}(\tau) \sim \frac{\eta}{8\pi\Gamma\tau}e^{4\sqrt{\Gamma\tau}}, 
\ee
where we have assumed a spatially independent seeding i.e., $\psi_{+1}(\xi,0)=\sqrt{\eta}$. 
Clearly, as a result of spatial propagation effects, both models predict a sub-exponential 
growth of the side-mode population. In the quantum model, however, the population grows like 
$e^{4\sqrt{\tau}}/\sqrt{\tau}$, whereas in the MF model it increases as 
$e^{4\sqrt{\tau}}/\tau$. The crucial point is that we cannot compensate for such a 
difference by assigning any constant value to the seeding $\eta$. Furthermore, the fact 
that the seeding appears as a prefactor in Eq. (\ref{forwg1_pfs}), suggests that any 
deviations of the MF atomic density profiles and populations from their quantum counterparts  
have to be attributed to the semiclassical nature of the MF model and not to 
the arbitrariness of the initial seeding, which may only lead to global changes such as  
a rescaling of the plotted curves.  
Keeping this in mind, we turn to comparing the predictions of the two models taking into account 
both forward and backward recoiling atoms, as well as a Thomas-Fermi distributed BEC. For direct comparison to previous work \cite{ZobNikPRA}, 
we have decided to seed the mean-field model according to  
\begin{equation}
\psi_{+1}(\xi,0)=\psi_0(\xi)/\sqrt{N},\label{seeding}
\end{equation}
which corresponds to one atom in the forward atomic sidemode distributed proportionally 
to the density of the BEC. 

\begin{figure}
\includegraphics[]{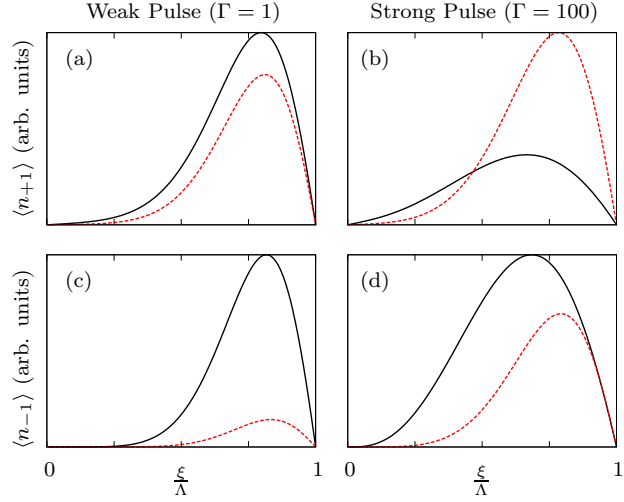}
\caption{(Color online) Atomic density profiles of the two side modes, according to the quantum model (black, solid) 
and the MF model (red, dashed), at $\Gamma\tau=6$. The left column shows the weak pulse regime and the right column 
the strong pulse regime.}\label{profiles}
\end{figure}

\begin{figure}
\includegraphics{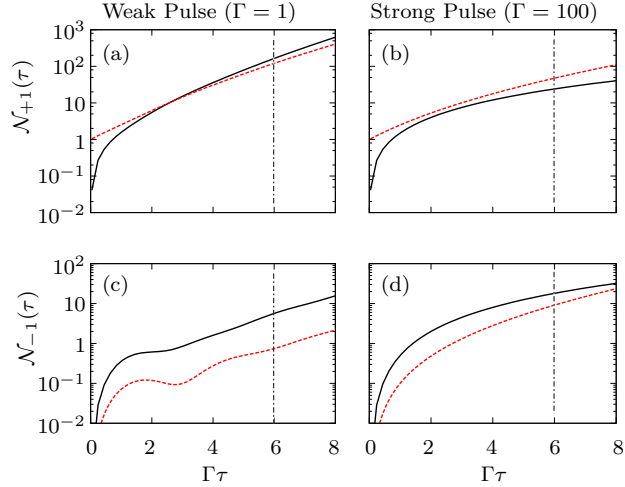}
\caption{(Color online) Evolution of the side-mode populations in the quantum (black, solid) and the MF (red, dashed) model.  Note that the quantities are plotted on a log-scale. The dot-dashed line marks the time when the snapshots in Fig. \ref{profiles} were taken.}\label{populations}
\end{figure}

A snapshot of the atomic density profiles in the two models, after a time $\Gamma\tau=6$,  
is plotted in Fig. \ref{profiles}. 
Both models predict that the profiles are peaked close to the right end of the condensate;  
a feature which is responsible for the experimentally observed asymmetry of the scattering 
pattern \cite{ZobNikPRA}. In the MF model, however, the profiles are peaked slightly 
closer to the end of the BEC, while the height and the width of the 
spatial distributions are underestimated, especially for the 
$(-)$ mode in the weak pulse regime. 
This discrepancy in the predictions of the two models is also reflected in the 
time evolution of the atomic populations in the two side modes, 
which are depicted in Fig. \ref{populations}. 

In agreement with experimental observations, both models predict that the populations of the 
two side modes are comparable in the strong-pulse regime, whereas in the weak-pulse regime, 
we have far less backwards than forwards recoiling atoms. This behavior is mathematically mirrored in the expressions for the atomic densities in the two side modes [see Eq. (\ref{densities})]. They differ by one 
term only, which is proportional to $\Gamma$, whereas their common term scales with $\Gamma^2$. 
Thus, for short times and strong pulses where $\Gamma\sim 10^2$, the two expressions become 
comparable, whereas they are different in the weak pulse regime where $\Gamma\sim 1$. 
It is also worth pointing out here that, as depicted in  Fig. \ref{populations}, 
the MF model gives approximately the right growth rates as well as the right qualitative behavior. 
In the quantum model, however, the suppression of the $(-)$ mode is not as prominent as in 
the MF description, and this can be attributed to the ambiguity of the seeding and therefore 
the initialization of the process. 

\subsection{Scattered light}
More insight into the differences between the two models is obtained by also studying 
the behavior of the radiation field. In the quantum model, the photon density within the 
BEC volume  model is given by Eq. (\ref{pd}), which in view of Eq. (\ref{e+1}) yields
\be
\mathfrak{I}(\xi,\tau)=
\frac{\Gamma}{\chi}\int_0^{\xi}{\mathrm d}\xi'|\varphi(\xi')|^2|F_{1,0}(\gamma_{\xi,\xi'},\tau)|^2.
\label{photondens}
\ee
Even though this quantity cannot be measured, we will use it to compare the two models as 
it influences many measurable features of the process. The most easily measurable observable 
linked to the radiation field is the number of photons which have left the condensate up to 
time $\tau$. Assuming no distracting factors between BEC and detector as well as instantaneous 
photon propagation, this quantity is given by Eq. (\ref{photonpop_out}).
Finally, it is straightforward to find the analogous quantities in the MF model 
using $\mathfrak{I}(\xi,\tau)=|e_+(\xi,\tau)|^2$, and again 
we can directly compare the predictions of the two models. 

\begin{figure}
\includegraphics{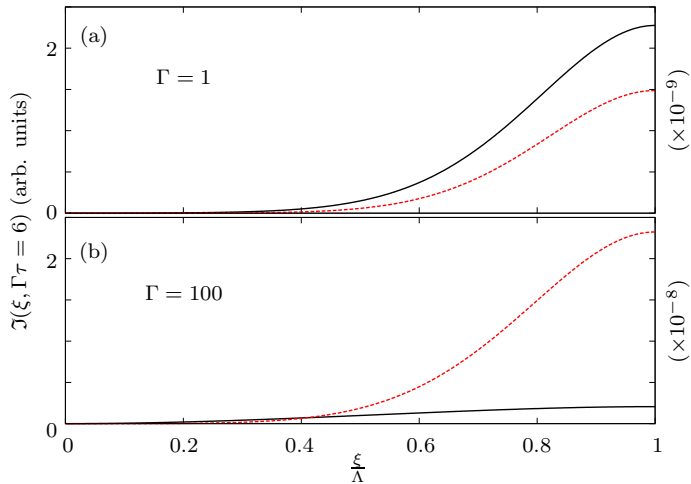}
\caption{(Color online) Comparison of the predictions for the photon density within the BEC in the quantum (black, solid) 
and the MF (red,dashes) model. The snapshots correspond to $\Gamma\tau=6$, in the weak-pulse ($\Gamma=1$) and the  
strong-pulse ($\Gamma=100$) regimes. }
\label{photondensfig}
\end{figure}

\begin{figure}
\includegraphics{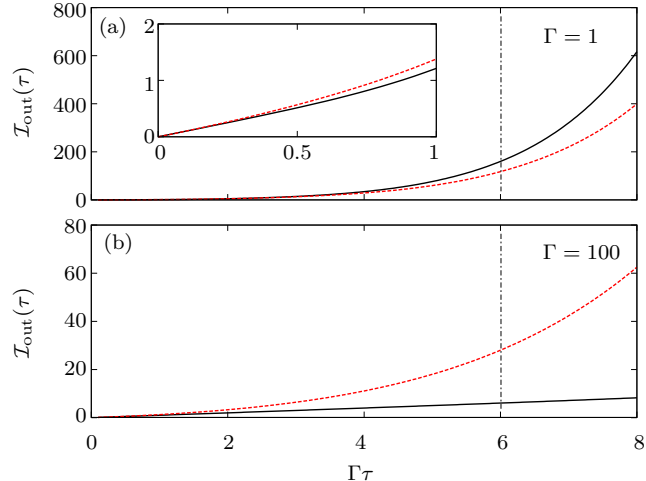}
\caption{(Color online) Comparison of predictions for the number of photons which have left the BEC as a function of scaled 
time $\Gamma\tau$ in the quantum (black, solid) and the MF (red, dashed) model. Top panel shows the weak-pulse regime 
($\Gamma=1$), while the lower panel is for the strong-pulse regime ($\Gamma=100$). 
The dot-dashed line marks the time when snapshots in Fig. \ref{photondensfig} are taken.
}
\label{photonpopfig}
\end{figure}

Calculations of the photon density within the BEC are plotted in Fig. \ref{photondensfig}.
In the weak pulse regime,  the mean-field model shows reasonable qualitative agreement with the quantum predictions. In the strong pulse regime, however, the quantum model predicts a much lower photon density than the MF model. This discrepancy becomes even more obvious if we look at the number of emitted photons, which is plotted in Fig. \ref{photonpopfig}. 
While the MF model predicts a fast growth of the number of photons, the growth in the quantum model is almost linear, which in view of Eq. (\ref{photonpop_out}) implies a constant density of photons within the BEC volume. Upon closer investigation, we find that such a period is also present in the weak pulse regime, albeit for shorter (scaled) times. More precisely, we find that for $\Gamma=1$ this period lasts only until about $\tau\approx 0.5$, as can be seen in the inset of fig \ref{photonpopfig} (a). 

According to our simulations, the presence of backwards recoiling atoms is suppressing 
superradiance. As depicted in Fig. \ref{photonpopfig}, the number of scattered photons with 
respect to scaled time in the weak pulse regime is much higher than in the strong pulse regime, since 
in the latter endfire photons are destroyed on account of producing 
backwards recoiling atoms. This removal inhibits the fast growth of the endfire mode, 
which in turn is responsible for the lower scattering rate (per scaled time). 
In particular, the endfire mode remains weakly populated for times $\tau\lesssim\tau_c$, since in this regime, 
the production of backwards recoiling atoms is allowed. This behavior is consistent with the conservation law (\ref{cont_eq}), which says that the number of photons outside the condensate is given by the number difference between the two matter-wave modes. 
Finally we note that the suppression of the population of the endfire mode seems to be underestimated in the MF model, 
as can be seen from Fig. \ref{photonpopfig} (b) as well as the inset of Fig. \ref{photonpopfig} (a).

\subsection{Averaging over semiclassical trajectories}
\label{rightseed}

\begin{figure}
\includegraphics{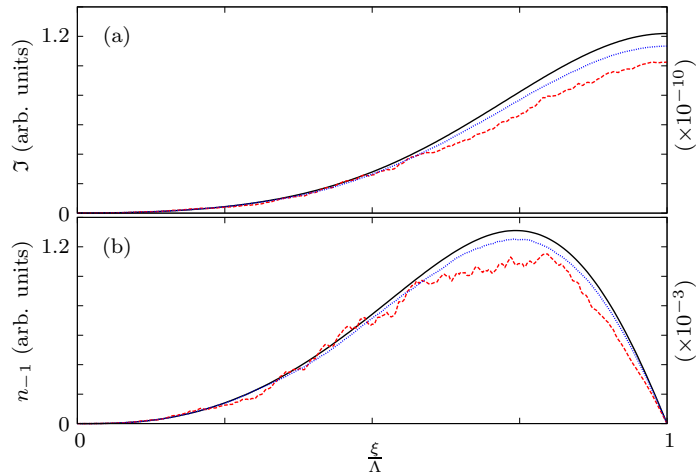}
\caption{
(Color online) Example of photon and atom densities obtained through averaging over randomly seeded MF solutions. 
Figure (a) shows a snapshot of the photon density and figure (b) shows the atomic density of the $(-1)$ mode. 
The quantum solution is shown in solid black, while averages over 20 and 2000 trajectories are shown in red dashed and 
blue dotted lines. Parameters: $\Gamma=1$, and $\tau=2$.}
\label{averagefig}
\end{figure}
For early times, the quantum prediction of the superradiant process is appropriate and can be expected to give better results than the MF model. It is easier, however, to perform calculations involving depletion of the BEC and population of higher modes in the MF model. It is therefore reasonable to ask, whether the quantum model is able to give us hints on how to seed the MF model appropriately to obtain quantitatively correct results. In the context of conventional superradiance, Haake {\it et al.} introduced the idea of averaging over many semiclassical ``trajectories'' to obtain quantum results \cite{Haake}. The MF equations are initially  seeded with random variables according to a particular distribution, while to obtain a particular quantum expectation value, one has to average over various solutions for the corresponding semiclassical quantity. We have investigated the extension of this idea to superradiant Rayleigh scattering off BECs where, in contrast to conventional superradiance,  backwards recoiling atoms are also present. 

To see how this works, let us assume we want to calculate the normal-ordered $n$th order correlation function 
\be
\langle\left[\hat\psi_{-1}^\dag(\xi_1,\tau)\right]^n\left[\hat\psi_{-1}(\xi_2,\tau)\right]^n\rangle.
\label{egcorrfunc}
\ee
From Eqs. (\ref{solutions}) and using (\ref{commrel}) as well as the fact that our initial state is the vacuum for all modes, we find the only non-vanishing expectation value involved to be 
\be
\langle\hat\psi_{+1}(\xi^{(1)},0)\ldots\hat\psi_{+1}(\xi^{(n)},0)\hat\psi_{+1}^\dag(\xi^{(n+1)},0)\ldots\hat\psi_{+1}^\dag(\xi^{(2n)},0)\rangle.
\label{normprodexpvalue}
\ee
Here, the variables $\xi^{(j)}$ are integrated from $0$ to $\xi_1$ for $j=1,\ldots,n$ and from $0$ to $\xi_2$ for $j=n+1,\ldots,2n$. Using the commutation relations (\ref{commrel}), correlation (\ref{normprodexpvalue}) reads
\be
\sum_\pi\prod_{j=1}^n\delta(\xi^{(j)}-\xi^{(n+\pi(j))}),
\label{contractions}
\ee
where $\delta$ denotes the Dirac delta and the sum runs over all permutations $\pi$ of order $n$. 
Let us now seed the semiclassical model with
\begin{subequations}
\bea
&&\psi_{+1}(\xi,0)=C_\xi,\\
&&\psi_{-1}(\xi,0)=0,\\
&&e_{+}(\xi,0)=0,
\eea
\end{subequations}
where $C_\xi$ is a random,  normally distributed complex variable, with zero mean and variance 
$1/\sqrt{\Delta\xi}$, with $\Delta\xi$ the spatial step of a numerical implementation. 
The average of the product $\psi_{+1}(\xi^{(1)},0)\ldots\psi_{+1}(\xi^{(n)},0)\psi_{+1}^*(\xi^{(n+1)},\tau)\ldots\psi_{+1}^*(\xi^{(2n)},0)$ over many trajectories (seedings) will effectively converge towards a discretized version of Eq. (\ref{contractions}). 
Due to the formal equivalence of the semiclassical solutions to the quantum ones, the product 
$\psi_{-1}(\xi_1,\tau)^n\psi_{-1}(\xi_2,\tau)^n$ will consequently converge to the quantum expectation value 
(\ref{egcorrfunc}). To find correlation functions of other operators, other seedings have to be used, 
which can be found in an analogous way. Table \ref{tab:1} summarizes these relationships. 
\begin{table}
\begin{ruledtabular}
\begin{tabular}{|c|c|}
\hline
Quantum Operator & Seeded Fields \\
\hline
\hline
$\hat e_+(\xi,\tau)$ & $\psi_{+1}(\xi',0)$ \\
\hline
$\hat\psi_{+1}(\xi,\tau)$ & $\psi_{-1}'(\xi',0)$\\
&$e_+(0,\tau')$  \\
\hline
$\hat\psi_{-1}(\xi,\tau)$ & $\psi_{+1}(\xi',0)$\\
\hline
\end{tabular}
\end{ruledtabular}
\caption{Summary of relations between quantum mechanical expectation values and random initial seeds to the mean-field model.}\label{tab:1}
\end{table} 

As far as densities and populations are concerned, the convergence of the averaging procedure towards the quantum 
solution is fairly fast. For instance, as depicted in Fig. \ref{averagefig}, one typically has to average over a 
couple of thousand trajectories, to obtain well converged, smooth density profiles. We note that the curves lie generally below the quantum mechanical results. This is, however, a purely numerical effect, and it disappears when the number of trajectories increases. For a sufficiently large sample of trajectories, we hence also have true numerical convergence.
Clearly, the averaging procedure is also applicable to correlations of higher order. 
One can calculate any normal ordered correlation function of the system by averaging over a sufficiently 
large ensemble of MF trajectories, provided that the operators involved in the correlation function have the 
same seeding requirements. The convergence, however, becomes rapidly slower with every order added, 
such that higher order correlations will require a larger number of trajectories.

\section{Results Beyond the Mean Field Model}
\subsection{Population Ratio}
\begin{figure}
\includegraphics[]{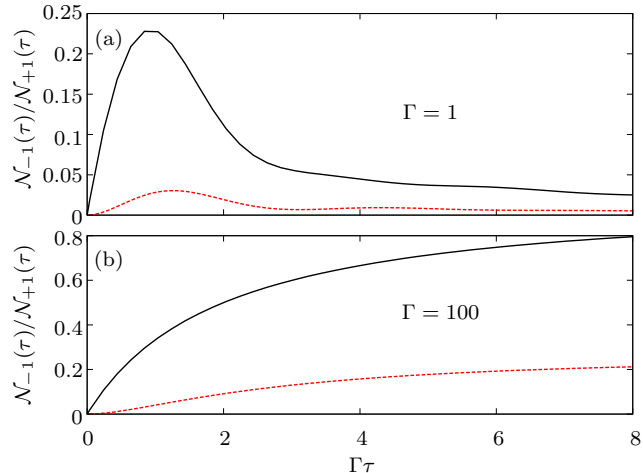}
\caption{(Color online)Time evolution of the ratio of backwards to forwards recoiling atoms as a function of scaled time 
in the quantum (black, solid) and the MF (red, dashed) model, for the weak-pulse (a) and the strong pulse (b) regimes.}
\label{ratioplot}
\end{figure}

An easily accessible quantity in a BEC superradiance experiment is the ratio of backwards to forwards recoiling atoms. In the quantum model, the calculation of this quantity is straight forward and unambiguous. From Eqs. (\ref{densities}) and (\ref{totalpops}) we find
\be
\frac{{\mathcal N}_{-1}(\tau)}{{\mathcal N}_{+1}(\tau)}=\left[1+\frac{\int_0^\Lambda{\mathrm d}\xi\int_0^\tau{\mathrm d}\tau'|\varphi(\xi)|^2|F_{1,0}(\gamma_{\xi,0},\tau')|^2}{\Gamma\int_0^\Lambda{\mathrm d}\xi\int_0^\xi{\mathrm d}\xi'|\varphi(\xi)|^2|\varphi(\xi')|^2|F_{1,1}(\gamma_{\xi,\xi'},\tau)|^2}\right]^{-1}.\label{popratio}
\ee 
One cannot expect the MF model to deliver reliable results for such a quantity, 
due to its ambiguous initialization related to the seeding. To find out how different the predictions of the two models for this ratio are, the MF equations of 
motion were seeded 
according to Eq. (\ref{seeding}), and the predictions of both models 
for the time evolution of the ratio in the two regimes are plotted in Fig. \ref{ratioplot}. 

Let us recall here that according to experimental observations, the population of the backwards atomic sidemodes is highly suppressed 
in the weak-pulse regime. 
Our quantum theoretical predictions reproduce these observations, i.e. the ratio is closer to one in the strong-pulse regime. 
In the MF model, it never exceeds $0.3$ throughout its evolution in either of the two regimes, even though it does attain higher values for strong couplings.   
As discussed in Sec. \ref{secII}, the suppression of backwards recoiling atoms due to the energy mismatch 
is expected to set in at $\tau\approx \tau_{c}$. Indeed, as depicted in Fig. \ref{ratioplot}, 
in the weak pulse regime the ratio of backwards to forwards recoiling 
atoms drops for times $\tau\gtrsim  \tau_{c}$. 
In the strong-pulse regime, however, and for the time scales consistent with the validity of 
our model, we are always well below $\tau_c$, and the ratio increases monotonically. 
Nevertheless, even in this case, the onset of the suppression manifests itself in the temporal 
behavior of the growth rate of the ratio.
Finally, it is also worth noting that the evolution of the ratio in the weak-pulse regime agrees qualitatively with the corresponding results in \cite{MooMeyPRL}. A quantitative comparison, however, is rather difficult due to different definitions of the coupling strength.

\subsection{Coherence of the scattered light}
Another class of questions typically addressed in a superradiance experiment pertains to the properties of 
emitted light. Of particular interest are the coherence properties which are described through correlation 
functions and are also accessible to measurements. For instance, 
the first-order correlation function $G^{(1)}(\tau,T)=\langle\hat{e}_+^\dag(\Lambda,\tau+T)\hat{e}_+(\Lambda,\tau)\rangle$, describes the temporal coherence properties of the light that are relevant to an 
experiment of the Young's type, where the light at two times (i.e., at $\tau$ and $\tau+T$) 
is superimposed to produce interference patterns \cite{textbook}. 
The visibility of the fringes in the observed pattern is proportional to 
the degree of first-order coherence, defined as  
\be
\label{g1_def}
g^{(1)}(\tau,\tau+T)=
\frac{G^{(1)}(\tau,\tau+T)}{\sqrt{\mathfrak I(\Lambda,\tau+T)\mathfrak I(\Lambda,\tau)}}.
\ee
On the other hand, intensity correlations are described through the 
normalized second order correlation function defined as \cite{textbook}
\be
g^{(2)}(\tau,\tau+T)=
\frac{\langle \hat{e}_+^\dag(\Lambda,\tau)\hat{e}_+^\dag(\Lambda,\tau+T)\hat{e}_+(\Lambda,\tau)\hat{e}_+(\Lambda,\tau+T) \rangle \rangle}{\mathfrak I(\Lambda,\tau) \mathfrak I(\Lambda,\tau+T)}.
\label{g2_def}
\ee
This quantity is basically related to the probability of detecting a photon at time $\tau+T$, 
given that a photon was detected at time $\tau$.  Definitions (\ref{g1_def}) and (\ref{g2_def}) are general and 
applicable to any light source, but can be simplified considerably 
for stationary sources, where the properties of the light depend only on the {\it delay time} $T$.  
Actually, this is the case typically discussed in standard text books \cite{textbook}. 
In the present setting, however, the process is by no means stationary, 
and thus well known expressions and conclusions are not necessarily  
applicable to our case. 

Using Eqs. (\ref{solutions}), one obtains 
\begin{subequations}
\label{chg2}
\be
g^{(2)}(\tau,\tau+T)=1+|g^{(1)}(\tau,\tau+T)|^2, 
\label{chaoticg2}
\ee
where 
\be
G^{(1)}(\tau,\tau+T)=\int_0^\Lambda {\mathrm d}\xi|\psi_0(\xi)|^2F_{1,0}(\gamma_{\Lambda,\xi},\tau)F_{1,0}^*(\gamma_{\Lambda,\xi},\tau+T),
\ee
\end{subequations}
while $\mathfrak I(\xi,\tau)$ is given by Eq. (\ref{photondens}).

\begin{figure}
\includegraphics[]{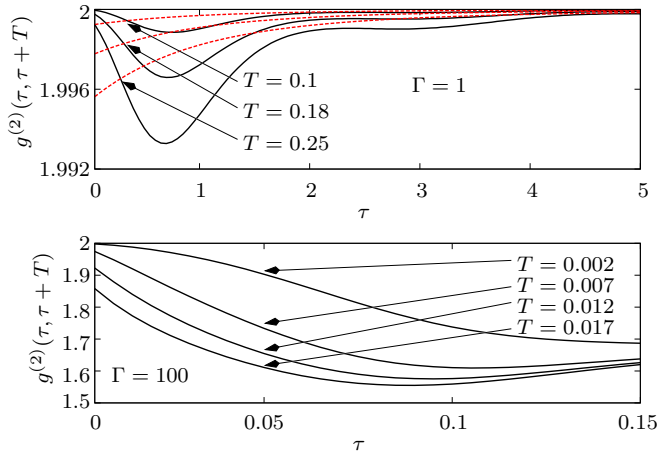}
\caption{(Color online) Behavior of the photon intensity correlation $g^{(2)}(\tau,\tau+T)$ as a function of $\tau$ for various fixed delays $T$ as given in the figures. The dashed line in the top figure gives the same function neglecting the $(-1)$ mode.}
\label{g2vstau}
\end{figure}

Equation (\ref{chaoticg2}) is a typical property of so-called chaotic light sources \cite{textbook}, 
albeit in our case the source is non stationary. Indeed, as depicted in Fig. \ref{g2vstau}, 
even within the undepleted pump approximation adopted throughout this work, 
for a given delay time $T$, $g^{(2)}(\tau,\tau+T)$ depends crucially on $\tau$. 
In view of Eq. (\ref{chg2}), we have $g^{(2)}(\tau,\tau)=2$ for all $\tau$,
which is a manifestation of intensity correlations. In other words, the endfire photons tend 
to appear bunched and thus detecting a photon at time $\tau$, significantly increases the 
probability of detecting another photon simultaneously. 
On the other hand, as $T\to\infty$ we obtain $g^{(2)}(\tau,\tau+T)\to 1$, indicating that the 
intensities are uncorrelated for large delay times. 
Typically, these asymptotic behaviors of  $g^{(2)}(\tau,\tau+T)$ are defined with respect to 
the characteristic  coherence time $T_c$ of the light under investigation (i.e., $T\to 0$ and 
$T\to\infty$ refer to $T\ll T_{c}$ and $T\gg T_{c}$, respectively). Unfortunately, 
the validity of the present model restricts us to relatively small times, and we cannot provide 
quantitative estimates of the coherence time $T_c$. Nevertheless, we can still draw some 
conclusions about the behavior of $g^{(2)}(\tau,\tau+T)$ in the weak and in 
the strong pulse regimes. Before this, it is also worth pointing out that measuring the 
correlation function 
in dependence of the delay time $T$ (irrespective of $\tau$) would facilitate any experiment 
considerably. 
In practice, this can be achieved, for instance, by forming blocks of data pertaining to various 
$\tau$ but the same delay time $T$, and then estimate $\tilde{g}^{(2)}(T)$ based  on these blocks. 
Formally, this procedure corresponds to the time averaged degree of second order coherence given 
by 
\be
\tilde{g}^{(2)}(T)=\frac{\int_0^{\tau_0}{\mathrm d}\tau\langle\hat{e}_+^\dag(\Lambda,\tau)\hat{e}_+^\dag(\Lambda,\tau+T)\hat{e}_+(\Lambda,\tau)\hat{e}_+(\Lambda,\tau+T) \rangle}{\int_0^{\tau_0}{\mathrm d}\tau\mathfrak I(\Lambda,\tau)\mathfrak I(\Lambda,\tau+T)}, \label{intg2}
\ee
where $\tau_0$ denotes the time over which experimental data are collected. This expression is analogous to volume integrated correlation functions used in \cite{NarGlau}. 
Note that when there is no dependence on $\tau$, Eq. (\ref{intg2}) reduces to the 
the standard expression of $\tilde{g}^{(2)}(T)$ for stationary sources \cite{textbook}.  

\begin{figure} 
\includegraphics[]{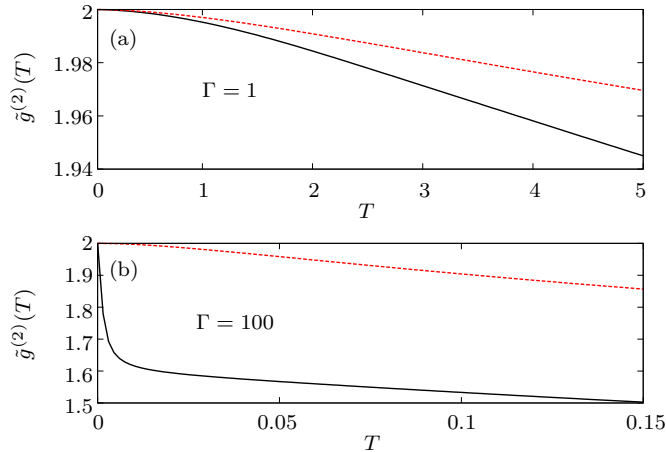}
\caption{(Color online) Integrated correlation function $\tilde{g}^{(2)}(T)$ as a function of the delay time $T$ for the weak (top) and strong (bottom) pulse regime. Dashed line in the top figure is the same function neglecting backwards recoiling atoms.}
\label{g2vsT}
\end{figure}

As depicted in Fig. \ref{g2vsT}, the behavior of $\tilde{g}^{(2)}(T)$ is substantially different 
in the weak and the strong pulse regimes. 
While in the weak pulse regime the correlation function seems to decay slowly but steadily, 
in the strong pulse regime we clearly have two stages. The initial transient regime is 
characterized by a rapid drop of $\tilde{g}^{(2)}(T)$, which is followed by a regime of 
very slow decay. To a good approximation $\tilde{g}^{(2)}(T)$ decreases 
linearly with increasing delay times, in both regimes. Moreover, according to Fig. \ref{g2vsT}, 
the tendency of photons to arrive in bunches is much lower in the strong pulse regime 
than in the weak pulse regime. 
This behavior can be attributed to the production of backwards recoiling atoms at the expense of 
endfire photons (see red curve). Intensity correlation function can thus viewed as a measure 
of the contribution of backwards recoiling atoms to the total number of scattering events. 

In this context, we can also interpret the behavior of $g^{(2)}(\tau,\tau+T)$ 
as a function of $\tau$. As depicted in Fig. \ref{g2vstau}, for both regimes 
there seems to be a systematic temporal behavior of $g^{(2)}(\tau,\tau+T)$ 
for a given delay time. The correlation function decreases for short times, while for larger 
times it increases (at least in the weak pulse regime). 
Such a behavior reflects changes in the statistical properties of the source, which can be 
also associated with the production of backward recoiling atoms. 
Indeed, as depicted in Fig. \ref{g2vstau}, neglecting the backward mode $(-)$ 
in our equations of motion, 
one finds a monotonic behavior of $g^{(2)}(\tau,\tau+T)$ (in good qualitative agreement with 
predictions for conventional superradiance \cite{Benedict}). For short times, photons that 
have been scattered into the endfire mode are consumed during the production 
of backward recoiling atoms which, as discussed in Sec. \ref{secII}, can take place for times shorter than 
$\tau_c$. This is also confirmed by the fact that according to Fig. \ref{g2vstau}, in the weak pulse regime 
for a given $T$, $g^{(2)}(\tau,\tau+T)$ exhibits a minimum for times very close to $\tau_c$. 
In the strong-pulse regime, although our time scales are always well below $\tau_c$,  
the onset of suppression manifests itself in the minimum of the intensity correlation 
function. To complete the picture, it is important to note here that superradiant Rayleigh 
scattering off BECs basically involves the mixing of optical and 
matter waves, which is a nonlinear process. Thus, any changes in the densities and/or populations of the 
fields that are mixed, are expected to affect considerably the statistics of the scattered light. 

\section{Conclusions}
 We have discussed the early stage of superradiant Rayleigh scattering off atomic condensates, in the framework of a quantum model that takes into account 
spatial effects.
Exact analytic solutions to this model can be expressed in terms of integrals involving Bessel functions, and are substantially different from the corresponding semiclassical solutions that have been obtained previously in the context of a related mean-field model. Nevertheless, the predictions of the two models about density profiles and growth rates are in reasonable qualitative agreement. An exception to this behavior is a strong suppression of photonic endfire modes at early times, which is underestimated in the mean-field treatment.
For a large collection of condensed atoms, the effect of quantum fluctuations 
on various observable quantities can be obtained by averaging over 
ensembles of semiclassical trajectories. Each trajectory corresponds to  
the solution of the mean-field equations of motion, where an appropriate random 
seeding has been used. Hence histograms of a particular observable will reveal its distribution according to the quantum model. This technique will be used in future work to study photon delay time statistics.
The quantum predictions for the ratio of backward to forward recoiling atoms is in qualitative agreement with experimental observations as well as other theoretical treatments.
Finally, the present model enabled us to calculate the statistical behavior of scattered photons, which is qualitatively different in the Kapitza-Dirac and Bragg regimes. This difference can be attributed to the suppression of photonic endfire modes due to the large number of back-scattered atoms at early times. In both regimes, the presence of backscattering distinguishes the photon statistics from the ones observed in conventional superradiance.
\section{Acknowledgement}
This work was supported by the EC RTN EMALI (Contract
No. MRTN-CT-2006-035369).
\begin{appendix}
\section{Solving the Equations of Motion}
To find solutions to Eqs. (\ref{eom}), we first apply the Laplace transform with respect to $\tau\to p$ to all three equations and find
\begin{subequations}
\begin{align}
\mathcal L_{\tau\to p}\{\hat\psi_{+1}^\dag(\xi,\tau)\}=&\frac{1}{p}\left[\hat\psi_{+1}^\dag(\xi,0)+\mathrm{i}\kappa\mathcal \psi_0^*(\xi)\mathcal L_{\tau\to p}\{\hat e_+(\xi,\tau)\}\right],\label{transmatfield1}\\
\mathcal L_{\tau\to p}\{\hat\psi_{-1}^\prime(\xi,\tau)\}=&\frac{1}{p+2\mathrm i}\left[\hat\psi_{-1}^\prime(\xi,0)-\mathrm{i}\kappa\mathcal \psi_0(\xi)\mathcal L_{\tau\to p}\{\hat e_+(\xi,\tau)\}\right],\label{transmatfield-1}\\
\chi\frac{\partial}{\partial\xi}\mathcal L_{\tau\to p}\{\hat e_{+}(\xi,\tau)\}=&-p\mathcal L_{\tau\to p}\{\hat e_{+}(\xi,\tau)\}+\hat e_+(\xi,0)\nonumber\\&-\mathrm{i}\kappa\left[\psi_0(\xi)\mathcal L_{\tau\to p}\{\hat\psi_{+1}^\dag(\xi,\tau)\}+\psi_0^*(\xi)\mathcal L_{\tau\to p}\{\hat \psi_{-1}^\prime(\xi,\tau)\}\right].\label{transmatfielde}
\end{align}
\end{subequations}
After inserting  (\ref{transmatfield1}) and (\ref{transmatfield-1}) in Eq. (\ref{transmatfielde}) we are left with a differential equation of the form
\be
\frac{\partial}{\partial\xi}\mathcal L_{\tau\to p}\{\hat e_{+}(\xi,\tau)\}=-\mathfrak{a}(\xi,p)\mathcal L_{\tau\to p}\{\hat e_{+}(\xi,\tau)\}-\mathfrak{b}(\xi,p),
\ee
which has the solution
\be
\mathcal L_{\tau\to p}\{\hat e_{+}(\xi,\tau)\}=e^{-\int_0^\xi\mathrm{d}\xi'\mathfrak{a}(\xi',p)}\mathcal L_{\tau\to p}\{\hat e_{+}(0,\tau)\}-e^{-\int_0^\xi\mathrm{d}\xi'\mathfrak{a}(\xi',p)}\int_0^\xi\mathrm{d}\xi' e^{-\int_0^\xi\mathrm{d}\xi''\mathfrak{a}(\xi'',p)}\mathfrak{b}(\xi',p).
\ee
This expression can be inserted in Eqs. (\ref{transmatfield1},\ref{transmatfield-1}), at which point we have closed expressions for all three Laplace transformed fields. The remaining inversion of the Laplace transforms can be found using elementary techniques given in e.g. \cite{abramovitz}.
The solutions are expressed in terms of the following functions
\begin{widetext}
\begin{subequations}
\begin{align}
F_{0,0}(y,z)=&
\delta(z)+\Theta(z)\sqrt{\frac{y}{z}}I_1\left(2\sqrt{yz}\right)
-\Theta(z)e^{-2\mathrm{i}z}\sqrt{\frac{y}{z}}J_1\left(2\sqrt{yz'}\right)\\
&-\Theta(z)y\int_0^z\mathrm{d}z'e^{-2\mathrm{i}z'}\frac{I_1\left(2\sqrt{y(z-z')}\right)J_1\left(2\sqrt{yz}\right)}{\sqrt{(z-z')z'}},\\
F_{1,0}(y,z)=&
I_0\left ( 2\sqrt{y z} \right )\Theta( z)
-\Theta( z)\sqrt{y}\int_0^{ z} dz^\prime \frac{e^{-{\rm
i}2z^\prime}}{\sqrt{z^\prime}}
I_0\left [ 2\sqrt{y(z-z^\prime)} \right ]
J_1\left ( 2\sqrt{y z^\prime} \right ),\\
F_{0,1}(y,z)=&
e^{-2{\rm i}z}J_0\left ( 2\sqrt{y z} \right )\Theta( z)+
\Theta( z)\sqrt{y}\int_0^{ z} dz^\prime \frac{e^{-{\rm
i}2z^\prime}}{\sqrt{z-z^\prime}}
I_1\left [ 2\sqrt{y(z-z^\prime)} \right ]
J_0\left (2\sqrt{yz^\prime} \right ),\\
F_{1,1}(y,z)=&\Theta( z)\int_0^{ z} dz^\prime
e^{-{\rm i}2z^\prime}
I_0\left [ 2\sqrt{y(z-z^\prime)} \right ]
J_0\left ( 2\sqrt{y z^\prime} \right ),\\
F_{2,0}(y,z)=&\sqrt{\frac{z}{y}}I_1(2\sqrt{yz})\Theta(z)-
\Theta(z)\int_0^z{\rm d}z^\prime e^{-2{\rm i}z^\prime}\sqrt{\frac{z-z^\prime}{z^\prime}}
I_1[2\sqrt{y(z-z^\prime)}]J_1[2\sqrt{yz^\prime}],
\\
F_{0,2}(y,z)=&e^{-2{\rm i}z}\sqrt{\frac{z}{y}}J_1(2\sqrt{yz})\Theta(z)+
\Theta(z)\int_0^{ z}{\rm d}z^\prime e^{-2{\rm i}z^\prime}\sqrt{\frac{z^\prime}{z-z^\prime}}
I_1[2\sqrt{y(z-z^\prime)}]J_1[2\sqrt{yz^\prime}],
\end{align}
\end{subequations}
\end{widetext}
with $J_i$ and $I_i$ the $i$th Bessel function of the first kind and the $i$th modified Bessel function respectively.
They satisfy the recursion relations
\begin{align}
\frac{\partial}{\partial u}F_{\mu,\nu}(u,v)=&F_{\mu+1,\nu}(u,v)-F_{\mu,\nu+1}(u,v),\nonumber\\
\frac{\partial}{\partial v}F_{\mu,\nu}(u,v)=&2F_{\mu-1,\nu}(u,v),\\
2{\textrm i}F_{\mu,\nu}(u,v)=&F_{\mu,\nu-1}(u,v)-F_{\mu-1,\nu}(u,v)\nonumber,
\end{align}
which are a consequence of properties of the Laplace transform. 
For the solutions with neglected backwards scattering, we use the functions
\bea
F_0(y,z)&=&\delta(z)+\Theta(z)\sqrt{\frac{y}{z}}I_1\left(2\sqrt{yz}\right)\\
F_1(y,z)&=&\Theta(z)I_0\left(2\sqrt{yz}\right)\\
F_2(y,z)&=&\Theta(z)\sqrt{\frac{z}{y}}I_1\left(2\sqrt{yz}\right).
\eea

\end{appendix}

\end{document}